\documentclass[twocolumn,amsmath,amssymb]{revtex4}
\usepackage{graphicx}% Include figure files
\usepackage{dcolumn}% Align table columns on decimal point
\usepackage{bm}% bold math
\newcommand{\etal}{\emph{et al.}}
\newcommand{\be}{\begin{equation}}
\newcommand{\ee}{\end{equation}}
\newcommand{\bfig}{\begin{figure}}
\newcommand{\efig}{\end{figure}}
\newcommand{\incl}{\includegraphics}
\begin{document}      

\title{Online Supplementary Material: \\
Low temperature vortex liquid in $\rm La_{2-x}Sr_xCuO_4$
} 
\author{Lu Li$^1$, J. G. Checkelsky$^1$, Seiki Komiya$^2$, 
Yoichi Ando$^2$, and N. P. Ong$^1$
}
\affiliation{
$^1$Department of Physics, Princeton University, Princeton, NJ 08544, USA\\
$^2$Central Research Institute of Electric Power Industry, Komae, Tokyo 201-8511, Japan
}

\date{\today}      
\begin{abstract}
\end{abstract}

\maketitle                   % Produces the title
\section{Introduction}
We report here the supplementary material for Ref. \cite{Li06}.  
Using torque magnetometry, we measured the magnetization anisotropy of 7 crystals of
$\rm La_{2-x}Sr_xCuO_4$ (LSCO), which are labeled as 03 (with $x$ = 0.030), 04 (0.040) 
05 (0.050), 055 (0.055), 06 (0.060), 07 (0.070) and 09 (0.090).

In optimally-doped cuprates, the bulk susceptibility is dominated by the paramagnetic van-Vleck orbital 
term $\chi_{orb}$ which has a significant anisotropy ($\chi^{orb}_c>\chi^{orb}_{ab}$) that 
changes weakly with $T$ (subscripts $c$ and ${ab}$ identify quantities measured with 
$\bf H||c$ and $\bf H\perp c$, respectively).   
Moreover, in the lightly-doped regime, the paramagnetic spin susceptibilities $\chi^s_c$ and $\chi^s_{ab}$ 
become significantly large below the interval 40-60 K.  However, the spin susceptibility is
very nearly isotropic (except below 10 K where its anisotropy becomes measurable).  
Against the large orbital and spin terms, the weak diamagnetic signal is very difficult to
resolve using standard bulk magnetometry in lightly-doped cuprates.  
By contrast, torque magnetometry selectively detects the orbital diamagnetism generated by
supercurrents confined to the CuO$_2$ layers while ignoring the large spin contribution when it is isotropic.  
The orbital van-Vleck contribution is also detected, but as a ``background'' that is 
$H$-linear to intense fields and only mildly $T$ dependent.  

\section{Experimental details}
\begin{figure}
\incl[width=9cm]{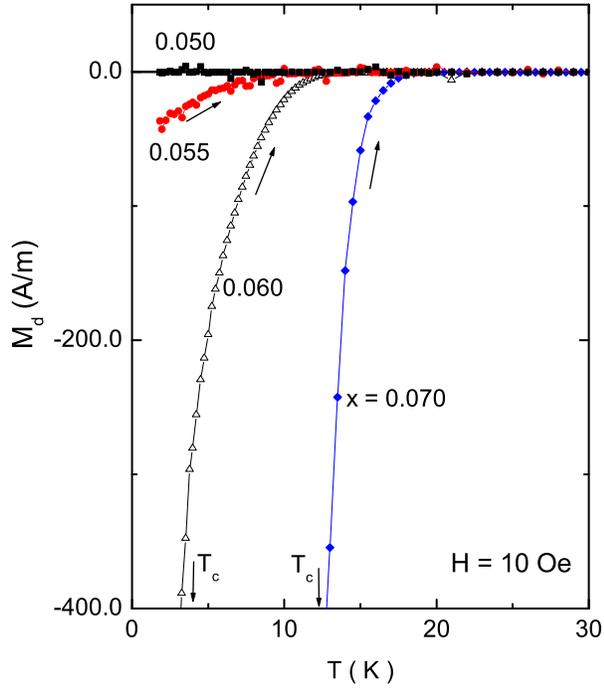}
\caption{\label{Meissner} 
The Meissner curves measured in weak field ($H$= 10 Oe) in samples 05, 055, 06 and 07 
measured after zero-field cooling.  The Meissner transitions are observed 
with mid-point temperature $T_c$ = 0, 0.5, 5, and 12 K in 05, 055, 06 and 07, respectively.
Full flux expulsion (with demagnetization factor $N_c\sim$ 2.2) corresponds to $M_d$ = -1,900 A/m.
It is instructive to compare these ZFC curves with the high-field curves in Fig. \ref{Mobs},
which are all in the reversible regime (except below 5 K in 07).
}
\end{figure}

Each crystal, of nominal size 2$\times$1$\times$0.35 mm$^3$, was glued to the tip of the cantilever with 
its $c$-axis at an angle $\phi \sim 15^{\mathrm o}$  
to $\bf H$.  The torque $\mbox{\boldmath{$\tau$}} = {\bf m\times B}$ leads 
to a flexing of the cantilever which is detected capacitively ($\bf m$ 
is the sample's magnetic moment and ${\bf B} = \mu_0({\bf H +M})$ 
with $\mu_0$ the vacuum permeability).  
We resolve $m\sim 5\times 10^{-9}$ emu at 10 T (1 emu = $10^{-3}$ Am$^2$).  
The torque measurements were
performed in-house in fields up to 14 T (down to $T$ = 0.35 K).  High-field measurements to
either 33 or 45 T were carried out at the National High Magnetic Field Laboratory (NHMFL), Tallahassee.   
Bulk measurements of all samples were performed to 2 K in a SQUID 
magnetometer (with resolution $\sim 10^{-6}$ emu) to calibrate the torque cantilever.
SQUID magnetometry was also used to observe flux expulsion by measuring the 
Meissner curves with $H$ = 10 Oe after zero-field cooling.  The curves in Fig. \ref{Meissner}
reveal full flux expulsion in sample 07, but only partial expulsion at the lowest $T$ (2 K) in 06 and 055.
No Meissner signal is observed in sample 05 down to 2 K.  The Meissner effect
requires the existence of long-range phase stiffness and coherence.

It is convenient to express the torque as an effective observed magnetization~\cite{Farrell,Bergemann}
$M_{obs}\equiv\tau/B_xV$, with $B_x = B\sin\phi$ (we take $\bf \hat{z}||\hat{c}$).
We have 
\be
M_{obs}(T,H_z) = M_d(T,H_z) + \Delta M_s(T,H_z) + \Delta\chi^{orb}(T)H_z,
\label{M3}
\ee
where $M_d$ is the diamagnetic term of interest.  Because the normal-state resistivity anisotropy
is extremely large in LSCO for $x<$0.10 ($\rho_c/\rho_{ab}$ = 6,000-8,000 below 40 K~\cite{Komiya}), 
we may assume that the supercurrents are predominantly in-plane.  Hence the in-plane component $M_{d,y}$
is negligible especially at fields above the melting field of the vortex solid $H_m$.
The term $\Delta M_s$ from the spin response is derived in Sec. \ref{spin}.
The anisotropy of the van Vleck orbital susceptibility $\Delta\chi^{orb}$ is the difference 
$\Delta\chi^{orb} = \chi^{orb}_c- \chi^{orb}_{ab}$ (hereafter, we write $H$ for $H_z$).  
In LSCO, $x_c\sim 0.055^{-}$.  

\begin{figure}
\incl[width=9cm]{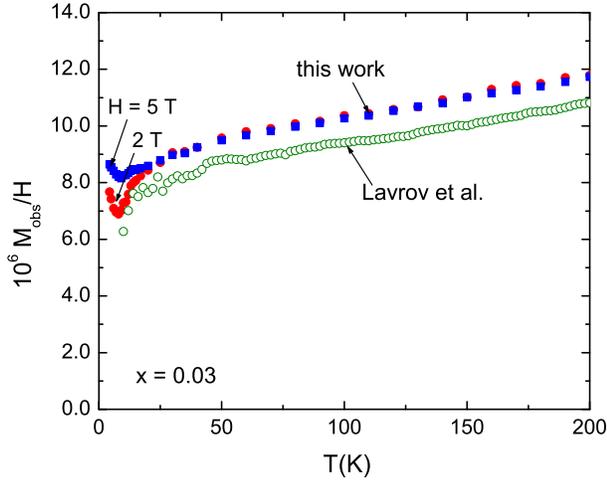}
\caption{\label{chiAndo} 
Comparison of the total $M_{obs}/H$ measured at $H$ = 2 and 5 T in sample 03
with the susceptibility anisotropy $\Delta\chi$ inferred from the bulk 
measurements of Lavrov \etal~\cite{Ando}.  In the latter, $\Delta\chi$ is defined 
as $\chi_c-\chi_{ab}$ where the in-plane susceptibility $\chi_{ab} = \frac12(\chi_a+\chi_b)$.  
The results of Lavrov \etal were measured by SQUID magnetometry
on a crystal with $x$ = 0.03 (and $\sim$10 times larger than 03 in volume) at 
the fixed field $H$ = 5 kOe.  Above 40 K, the agreement is quite good (the slight
offset may come from absolute calibration of the torque cantilever).  Below 40 K, the 
ratio $M_{obs}/H$ is strongly $H$ dependent because of increased contributions from
$M_d$ and $\Delta M_s$.
}
\end{figure}

As a check, we have compared the ``raw'' data $M_{obs}/H$ in sample 03 against the 
SQUID magnetometry measurements of Lavrov \etal~\cite{Ando} (Fig. \ref{chiAndo}).  They have reported
detailed measurements of the total bulk susceptibilities $\chi_a$, $\chi_b$ and $\chi_c$ 
in a very large LSCO crystal ($x$ = 0.03) with a volume $\sim$10 times 
larger than in our samples.  In Fig. \ref{chiAndo} we plot the $T$ dependence of the
anisotropy $\Delta \chi = \chi_c-\frac12(\chi_a +\chi_b)$.  There is good agreement 
with our corresponding quantity $M_{obs}/H$ measured at 2 and 5 T above 40 K.

\section{Magnetization results}
In the high-field experiments at the NHMFL, all samples (03--09) were investigated to 
a peak field of 33 T (using Bitter magnets).  Further measurements were done on Sample 06 
in the hybrid magnet in the field range 11-45 T.  In all samples, the high-field data were 
supplemented by extensive ``low-field'' in-house results (up to 14 T), which have improved
resolution below $\sim$5 T.

Figure \ref{Mobs} shows the general trend of the ``raw'' observed magnetization 
$M_{obs}$ vs. $H$ in Samples 03--07 using the low-field data sets.
Above the onset temperature for the appearance of vortex-Nernst signal $T_{onset}$,
($\sim$ 30, 40, 60 K in samples 04, 05 and 06, respectively), the
$H$-linear term $\Delta\chi^{orb}H$ dominates to produce a fan-like pattern.
Below $T_{onset}$, the rapid growth of $M_d(T,H)$ which is strongly nonlinear in $H$
produces a noticeable downward deviation from the fan pattern.  This is strikingly evident in 055, 06 and 07
but is also seen in 03, 04 and 05.  When $T$ decreases below $T_c$, the samples with $x>x_c$
(055, 06 and 07) display a large diamagnetism even in the limit $H\rightarrow 0$.
This requires the vortex solid to be stable.  In samples with $x<x_c$ (03--05), however, the
diamagnetic signal does not grow significantly even with cooling to 0.35 K.

\emph{Susceptibility}  The 3 distinct contributions (Eq. \ref{M3}) to the observed magnetization $M_{obs}$ 
are actually readily apparent in the raw magnetization data when the field is extended to 33 Tesla.
In a linear plot of $M_{obs}$ vs. $H$ (as in Fig. \ref{Mobs}a), the terms $M_d$ and $\Delta M_s$
tend to be dwarfed by the large van-Vleck term $\Delta\chi^{orb}H$, and hard to make out.
On the other hand, by simply dividing the raw magnetization by $H$ to form the observed 
susceptibility $\chi_{obs}\equiv M_{obs}/H$, we can make all 3 terms apparent.
Figure \ref{chi} shows the field dependence of $\chi_{obs}(T,H)$ in Samples 03--055 
at selected $T$ [we display $\chi_{obs}(T,H)$ divided by its curve at 40 K].  First, we look at 
Panel d (Sample 055).  Cooling from 80 K to 35 K, we observe that $\chi_{obs}$ is $H$
independent with a weak $T$ dependence.  This high-$T$ behavior of $\chi_{obs}$ 
leads to the ``fan-like'' pattern identified with the van-Vleck magnetization $\Delta\chi^{orb}H$ in Fig. \ref{Mobs}.
The negative contribution of $M_d/H$, first resolved near 50 K, grows rapidly in 
magnitude as $T$ decreases to 5 K.   Moreover, while $M_d/H$ is very large in low fields, it is 
suppressed when $H$ exceeds 33 T, consistent with field suppression of the pair condensate.
Sample 05 in Panel c shows a similar pattern of behavior
except that, below 10 K, a new positive, strongly $H$-dependent contribution emerges to lift
the curves upwards (compare curves at 5 and 10 K).  Proceeding to Samples 04 and 03 (Panels
b and a, respectively), we see that the new paramagnetic term becomes steadily larger with deceasing $x$.  
This term -- associated with the spin contribution $\Delta M_s/H$ -- is clearly distinguishable from 
the van-Vleck and diamagnetic terms.  In spite of the large spin term in 03 and 04, 
the diamagnetic term $M_d/H$ remains quite robust up to fields of 20-30 T.

\emph{Removal of van Vleck term}
As explained in the main text~\cite{Li06}, it is best to subtract altogether the large ``background''
van Vleck term $\Delta\chi^{orb}$ to analyze the diamagnetic and spin terms accurately.
The broad interval of $T$ in the low-field data set allows $\Delta\chi^{orb}(T)$ to be measured accurately.  
As an example, Fig. \ref{onset} displays the $T$ dependence of $M_{obs}$ (measured at 5 T) 
in sample 06.  When $T>$ 50 K, $M_{obs}$ is dominated by $\Delta\chi^{orb}H$ which has
the weak $T$-linear dependence
\be
\Delta\chi^{orb}(T) = A\left(1+\frac{T}{T_0} \right)
\label{orb}
\ee
with $A\sim$ 1.05$\times 10^{-5}$ and $T_0\sim$ 700 K (dashed line).  The temperature scale $T_0$
is independent of $x$ to our resolution, while the intercept parameter $A$ shows a very
weak $x$ dependence (both are associated with the van Vleck background).  
However, $T_{onset}$ decreases significantly, nominally linearly
with $x$, as $x$ falls below 0.07.  
The change between 0 and 40 K of this background term is $<6\%$.
Using Eq. \ref{orb}, we remove the van Vleck term to obtain the curves of $M_{orb}'(T,H)$,
as explained in Ref. \cite{Li06}.

As anticipated in Fig. \ref{chi}, the background-subtracted curves $M_{orb}'$ 
reveal an interesting competition between the diamagnetic
term $M_d(T,H)$ and the spin term $\Delta M_s(T,H)$ as $x$ changes from 0.03 to 0.06 (Fig. \ref{MvsH}).
In sample 03, $M_d$ though small is still strong enough to pull $M_{obs}'$ negative 
below $\sim$10 T.  Above 10 T, $M_d$ is rapidly suppressed to zero at 24 T, leaving the spin term
which is nominally $H$ independent.  Turning to Panel d for sample 06, we see the same competition
between the 2 terms, except that $M_d$ is now much larger, and a larger depairing field ($\sim$48 T) is 
needed to suppress it to zero.  Also the saturation value of $\Delta M_s$ is only $\frac12$ as large.
The behaviors in 04, 05 and 055 are intermediate between these 2 extremes.

\emph{Paramagnetic spin-$\frac12$ moments}
The overall behavior of the spin term in high fields, together with the oscillatory behavior
in weak $H$ (see Fig. 1c of Ref. \cite{Li06}) suggests that the spin term arises from 
spin-$\frac12$ local moments that are nearly non-interacting but have 
anisotropic $g$-factors ($g_{ab}<g_c$) that are weakly $T$-dependent.
As derived in Sec. \ref{spin}, the spin contribution to $M_{obs}$ is
\be
\Delta M_s(T,H) = {\cal P}(T)\tanh[\beta g_{\phi}\mu_BB/2],  
\label{Ms}
\ee
with $\mu_B$ the Bohr magneton and $\beta = 1/k_BT$.  The prefactor
is ($N_s$ is the spin population)
\be
{\cal P}(T) = \frac{N_s\mu_B}{V} \frac{(g_c^2 - g_{ab}^2)}{2g_{\phi}} \cos\phi.
\label{P}
\ee
with the effective $g$-factor
\be
g_{\phi} = \sqrt{(g_c\cos\phi)^2+(g_{ab}\sin\phi)^2}.
\label{g}
\ee
With $g_{\phi}\sim g_c$ fixed at 2.1, the only adjustable parameter at each $T$ is ${\cal P}$.

Using Eq. \ref{Ms}, we may subtract the contribution of the spin term from $M_{obs}'$ to 
obtain the diamagnetic term $M_d$ at each $T$.  The curves of both $\Delta M_s$ and $M_d$
in sample 03 are displayed in Fig. \ref{M03}.  The profile of $M_d$ resembles the 
``tilted-hill'' profile observed for the vortex-Nernst signal observed at $T>T_c$ in
both underdoped LSCO (at large doping) and in Bi-based cuprates.  The curves here are 
all associated with the vortex liquid (the vortex solid is not observed in 03 down to 0.35 K).

We remark that the diamagnetic term $M_d$ is readily apparent in the 
raw data of $\chi_{obs}$ vs. $H$ in all samples.  Moreover, it remains robust up to 
high fields of 25-45 T (Fig. \ref{chi}), so the evidence for 
the vortex-liquid diamagnetism in samples 03--06 does not depend on our background subtraction
or the form of Eq. \ref{Ms}.  The merit of Eq. \ref{Ms} is that it allows us to understand the details
of the low-$T$ oscillatory behavior of $M_{obs}'$ shown in Fig. 1c of Ref. \cite{Li06}, and to extract
$M_d(T,H)$ with good accuracy.

In Fig. \ref{M03}, the paramagnetic curves are the best fits using Eq. \ref{Ms} with 
the sole adjustable parameter ${\cal P}(T)$ which is plotted in the inset.  
Below $T_{sg}\sim$ 2.5 K in this sample, the curves of $M_{obs}'$ vs. $H$ become 
slightly hysteretic reflecting the onset of spin-glass behavior.  The spin-glass hysteresis loops
are clockwise in contrast with the anti-clockwise sense of the loops in the 
vortex solid (we will discuss the spin-glass observations elsewhere).

\section{Anisotropic spin term}\label{spin}
The Hamiltonian for an $s = \frac12$ spin with anisotropic $g$-factor in a 
field $\bf B$ (lying in the $x$-$z$ plane at an angle $\phi$ to $\bf z ||c$) is 
\be 
H = -\frac12\mu_B B(g_{ab}\sigma_x \sin\phi + g_c\sigma_z \cos\phi),
\label{Hspin}
\ee
where $\sigma_x$ and $\sigma_z$ are the Pauli matrices and $B={\bf |B|}$.  Because of
the anisotropy, the tilt-angle $\theta$ of $\vec{\sigma}$ differs from $\phi$.
The eigen-spinors of Eq. \ref{Hspin} are
\be
|+\rangle = \left( \begin{array}{c}
\cos\frac{\theta}{2} \\
\sin\frac{\theta}{2}
\end{array}\right),  \quad\quad
|-\rangle = \left( \begin{array}{c}
-\sin\frac{\theta}{2} \\
\cos\frac{\theta}{2}
\end{array}\right)
\label{spinor}
\ee
with $\theta$ defined by
\be
\cos\theta = \frac{g_c}{g_{\phi}}\cos\phi, \quad \sin\theta = \frac{g_{ab}}{g_{\phi}}\sin\phi,    
\label{theta}
\ee
and $g_{\phi}$ as given in Eq. \ref{g}.
With Eq. \ref{spinor}, the matrix elements are $\langle +|\sigma_z|+\rangle = \cos\theta$ 
and $\langle +|\sigma_x|+\rangle = \sin\theta$ etc. 

For a collection of $N_s$ spins in a volume $V$ in a bath at temperature $T$, 
the thermal-averaged magnetization is 
\be
\langle {\bf M}_s\rangle = \frac{N_s\mu_B}{2V} [g_c\cos\theta {\bf\hat{z}} + 
g_{ab}\sin\theta{\bf \hat{x}}] \tanh[\beta g_{\phi}\mu_BB/2].
\label{M}
\ee
This contributes the torque signal
\be
\tau_{s} = N_s \mu_B  \frac{(g_c^2 - g_{ab}^2)}{2g_{\phi}}B \cos\phi\sin\phi \tanh[\beta g_{\phi}\mu_BB/2],
\label{tau}
\ee
from which we obtain the results in Eqs. \ref{Ms}--\ref{g}.

\section{Hysteretic curves and vortex avalanches}
\begin{figure*}
\incl[width=15cm]{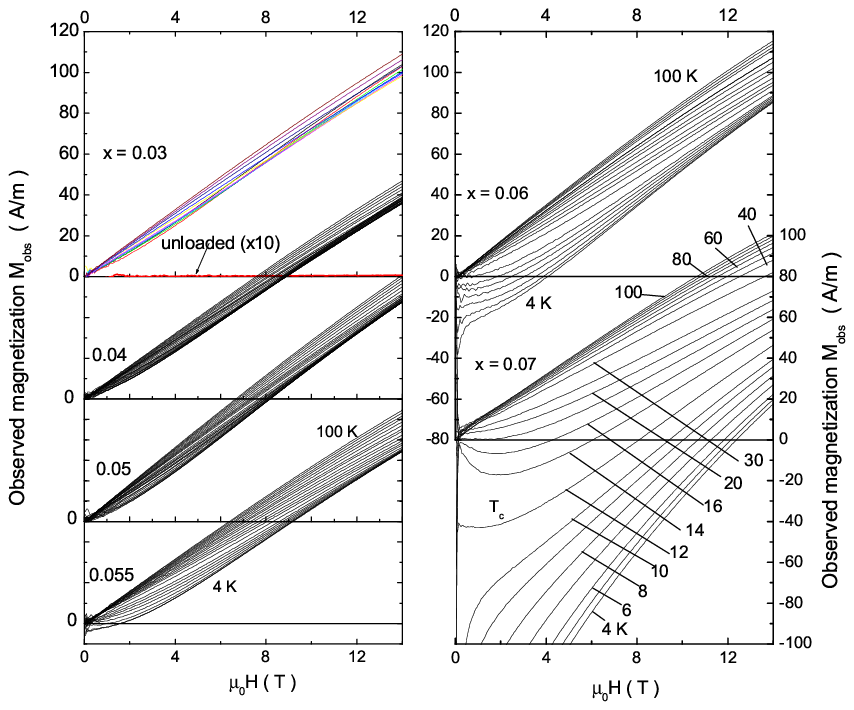}
\caption{\label{Mobs} 
The total (observed) magnetization $M_{obs}$ in the 6 LSCO crystals 03, 04, 05, 055, 06
and 07 with doping $x$ = 0.03, 0.04, 0.05, 0.055, 0.06 and 0.07, respectively.  The Meissner
transition midpoint $T_c \sim$ 0.5, 5 and 12 K in 055, 06 and 07.  The range of 
$T$ in each sample (4-100 K) are similar to the ones indicated for sample 07.  All curves
share the same scales for the $M_{obs}$ and $H$ axes, but the origins have been 
shifted vertically for clarity.  The background paramagnetic term $\Delta\chi^{orb}H$ 
is strictly $H$-linear and weakly $T$ dependent (nominally the same in all samples).  
Below 30 K, the growth of the diamagnetic term $M_d$,
and the spin anisotropy $\Delta M_s$, become increasing evident from 03 to 07.  The horizontal bold
curve plotted with sample 03 shows the signal of the unloaded cantilever amplified by 10.
All curves are in the reversible regime, except in sample 07 where hysteresis appears
below 5 K.  See Ref. \cite{Li06} for discussion of irreversibility.
}
\end{figure*}

\begin{figure*}
\incl[width=14cm]{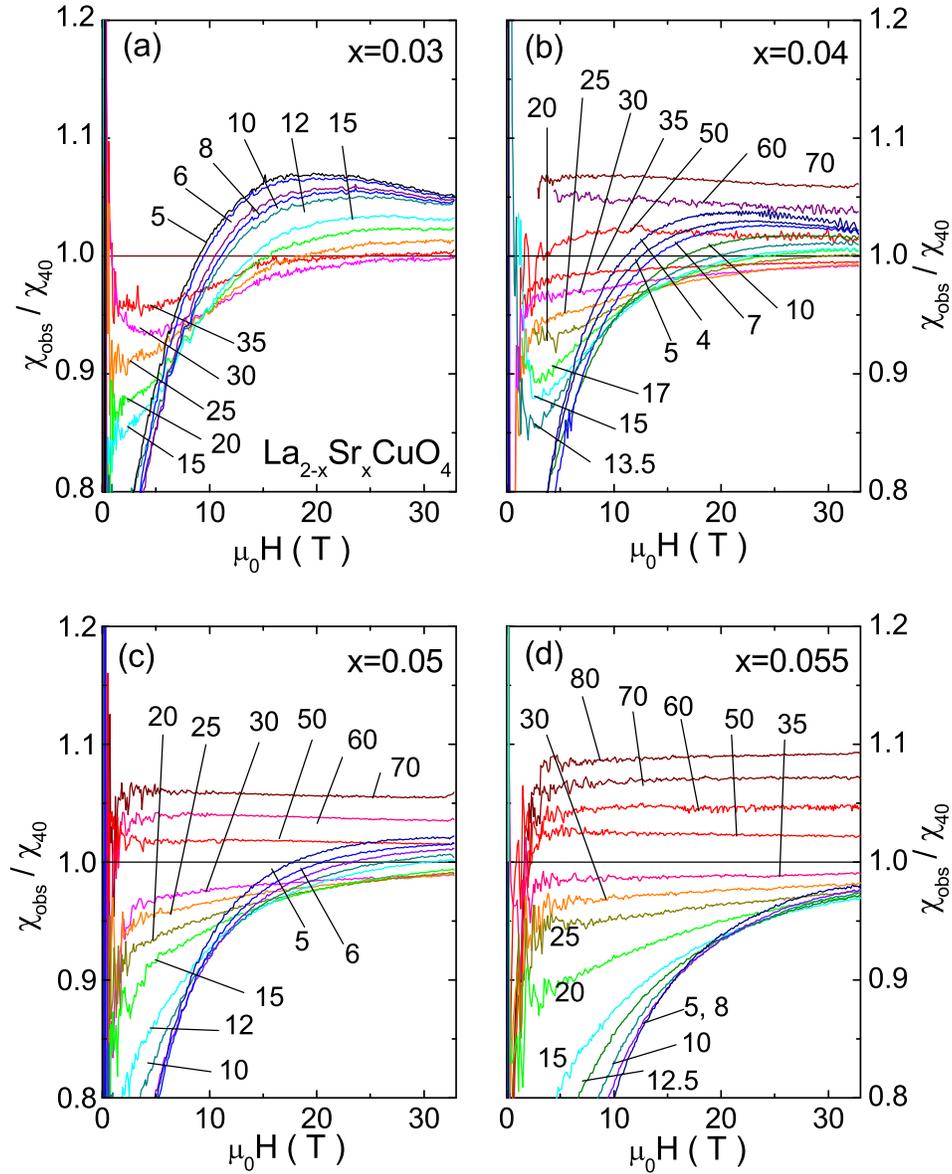}
\caption{\label{chi}  The $H$ dependence of the observed susceptibility $\chi_{obs}\equiv M_{obs}/H$ 
at selected $T$ in Samples 03, 04, 05 and 055 (Panels a--d, respectively).  $\chi_{obs}(T,H)$ is plotted divided by
its value at 40 K.  In each panel $\chi_{obs}$ is $H$ independent and only weakly $T$ dependent above $\sim$35 K
consistent with the van-Vleck term.  The strongly $H$-dependent diamagnetic term $M_d/H$ grows rapidly
between 20 and 5 K.  Finally, at low $T$, a third positive contribution -- the paramagnetic 
spin term $\Delta M_s/H$ -- emerges to counter the diamagnetic term in intense fields.  At 5 K, the spin term is barely
visible in 055, but quite large in Samples 03 and 04.
}
\end{figure*}

\begin{figure}
\incl[width=8cm]{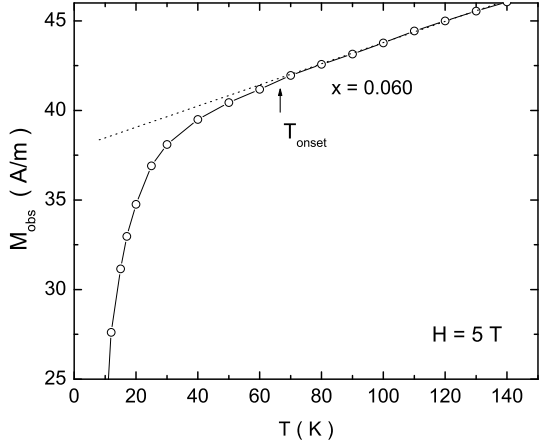}
\caption{\label{onset}
The $T$ dependence of the total observed torque magnetization $M_{obs}$ measured 
at $H$ = 5 T in sample 06
showing the weak $T$-linear change above $T_{onset}$ (arrows), the onset temperature of diamagnetism.
The linear extrapolation (dotted lines) below $T_{onset}$ is used to determine the anisotropic van Vleck
contribution $\Delta\chi^{orb}H$.  Below $T_{onset}$, the magnitude $|M_d|$ increases rapidly.  
}
\end{figure}

\begin{figure*}
\incl[width=14cm]{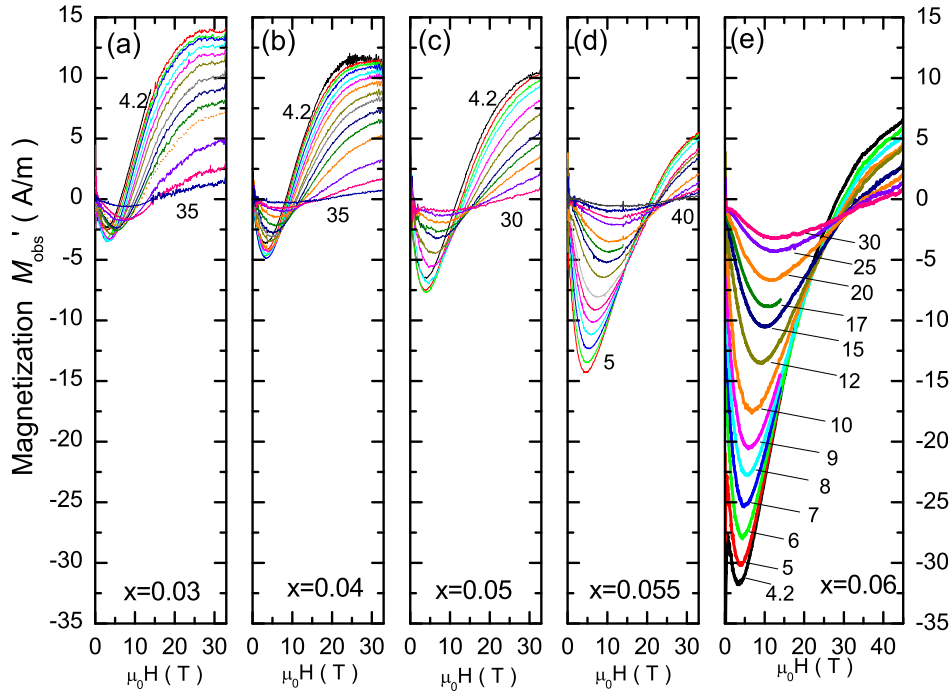}
\caption{\label{MvsH} 
The curves of $M_{orb}'$ (observed magnetization with van Vleck term removed)
in samples 03, 04, 05, 055 and 06 (Panels a, b, c, d and e, respectively).  Each of these curves
is comprised of a diamagnetic term $M_d(T,H)$ that is dominant in low $H$ ($<$10 T),
but is suppressed above 20-30 T, and a paramagnetic spin term $\Delta M_s(T,H)$ 
that is nearly constant in $H$ in high fields.  As we increase $x$, the diamagnetic term
$M_d$ becomes increasingly dominant and harder to suppress in high fields (note expanded 
field scale in Panel d).  Samples 055 and 06 show broad Meissner transitions with $T_c\sim$ 0.5
and 5 K, respectively.  The other samples show no Meissner transitions.  For clarity,
we have not displayed curves at $T$ below 4 K (see Fig. 2 in \cite{Li06}).
}
\end{figure*}

\begin{figure}
\incl[width=8cm]{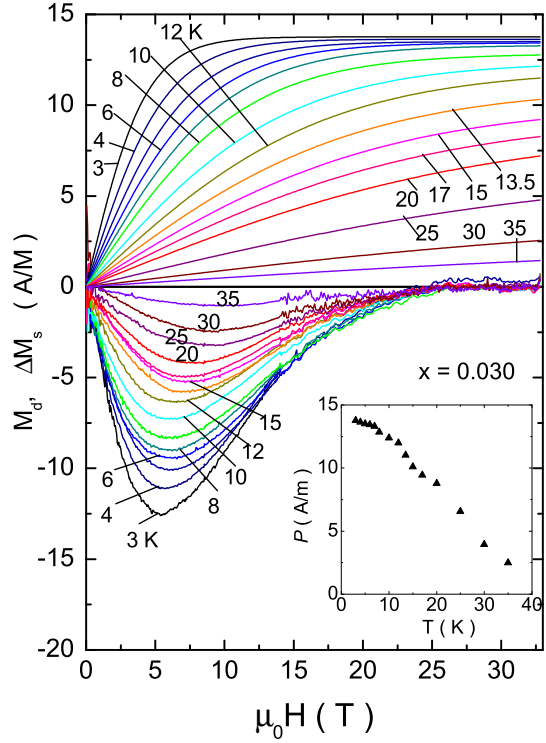}
\caption{\label{M03}  
Curves of the the diamagnetic term $M_d(T,H)$ and the paramagnetic spin term $\Delta M_s(T,H)$ 
in sample 03 ($x$ = 0.03) at fixed $T$ as indicated.  
At each $T$, after removing the orbital van-Vleck term, the remaining signal is given by
$M_{obs}'(T,H) = \Delta M_s(T,H) + M_d(T,H)$.  The plotted curves of $M_d$ and
$\Delta M_s$ add to reproduce $M_{obs}'$.  The inset shows the $T$ dependence
of the prefactor ${\cal P}(T)$ obtained from the fit to Eq. \ref{Ms}.
}
\end{figure}

\begin{figure*}
\incl[width=14cm]{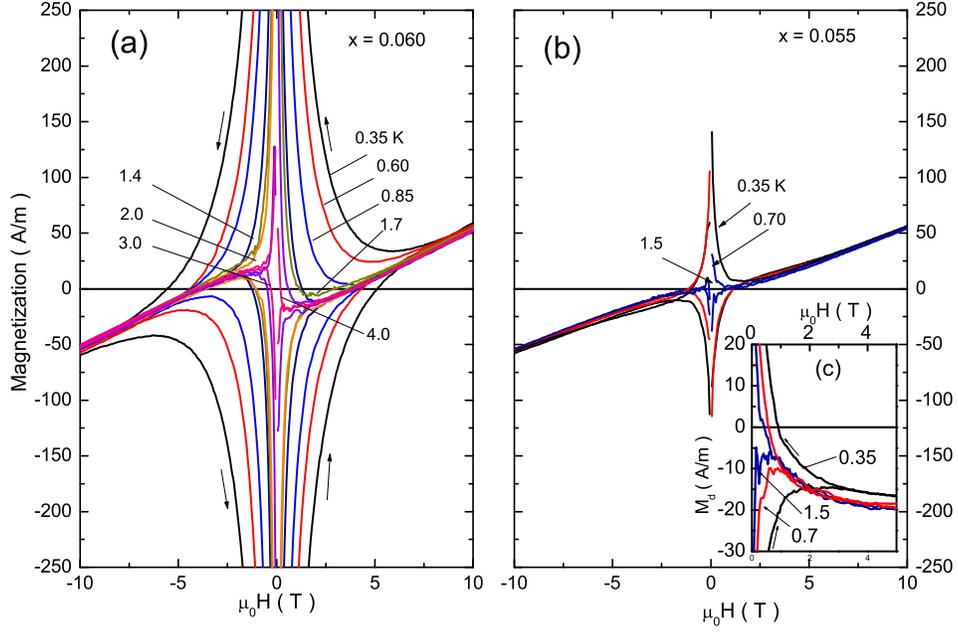}
\caption{\label{Hyst} Comparison of the hysteretic loops in Samples 06 (Panel a) and 055 (b).
In spite of the small difference in $x$, the hysteresis loops are much larger in 06 than in 055
measured at the same $T$.  This reflects the rapid decrease of the zero-Kelvin irreversibility
field $H_{irr}(0)$ as $x$ approaches $x_c\sim$ 0.055.  The inset (Panel c) shows an expanded view
of the hysteresis at 0.35, 0.7 and 1.5 K in Sample 055.
}
\end{figure*}

A way to see the rapid suppression of the vortex solid state as $x\rightarrow x_c^{+}$
is to compare the hysteretic curves of Samples 06 and 055 at low temperatures (Fig. \ref{Hyst}).
In going from 06 to 055 at $T$ = 0.35 K, the width of the hysteretic loops at a fixed field ($H$ = 1 T) 
shrinks by $\sim$100.  Between 07 and 06, the loops shrink by another factor of 10 (see Fig. \ref{Mobs}).  The
1000-fold decrease between 07 and 055 provides evidence for a critical transition at $x_c$.  The opposing view
of a distribution of superconducting islands implies a gradual decrease of the
hysteretic loops that is inconsistent with our observations.

\begin{figure}
\incl[width=8cm]{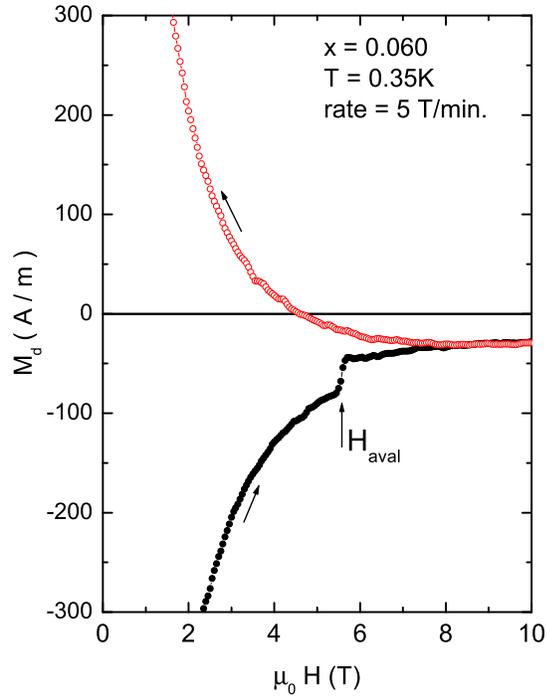}
\caption{\label{jump}
Jumps in the field sweep-up portion of the hysteretic curves 
of $M_d$ vs. $H$  (arrow at $H_{aval}$) in sample 06 at $T$ = 0.35 K at 
a field sweep rate of 5 T/min.  The jumps are not observed when the
sweep rates are reduced by a factor of 10. Vortex avalanches are observed only in the
vortex-solid phase.
}
\end{figure}

At $T$ = 0.35 K, vortex avalanches can be triggered 
in the vortex solid ($H<H_{irr}(T)$) by sweeping the field at a rate of 3--5 T/min.
As shown in Fig. \ref{jump}, the jumps occur during the field sweep-up portion of the
hysteretic loops.  By decreasing the sweep rate by a factor of 10, we can 
eliminate the jumps altogether.  Vortex avalanches occur when the field sweep is too
rapid to allow the inserted vortices to equilibrate in the solid phase.  They involve collective
depinning of a large fraction of the vortices in the crystal in a very short time interval, and
are generally considered to be direct evidence for the existence of the vortex solid.

The research at Princeton 
was supported by the National Science Foundation (NSF) through a MRSEC grant DMR 0213706.
Research at CRIEPI was supported by a Grant-in-Aid for Science from the Japan Society for 
the Promotion of Science.  The high field measurements were performed in the 
National High Magnetic Field Lab. Tallahassee, which is supported by NSF, the 
Department of Energy and the State of Florida.

\newpage

%%%%%%%%%%%%%%%%%%%%%%%%% FIG 1
%%%%%%%%%%%%%%%%%%%%%%%%

%%%%%%%%%%%%%%FIG 2

%%%%%%%%%%%%%%%%% FIG 3
%%%%%%%%%%%%%%%%%%%%%%%%

%%%%%%%%%%%%%%%%% FIG 4

%%%%%%%%%%%%%%%%%%%%%%%%%% FIG 5

%%%%%%%%%%%%%%%%%%% FIG 6
%%%%%%%%%%%%%%%%%%

%%%%%%%%%%%%%%%%%%%%% FIG 7
%%%%%%%%%%%%%%%%%%%%%%%%

%%%%%%%%%%%%%%%  FIG 8

%%%%%%%%%%%%%%%%%%%%% FIG 9
\end{document}